\documentclass[a4paper]{jpconf}
\usepackage{graphicx}
\begin{document}
\title{Superconducting density of states at the border of an amorphous thin film grown by focused-ion-beam}

\author{I. Guillam\'on$^1$, H. Suderow$^1$, S. Vieira$^1$, A. Fern\'andez-Pacheco$^{2,3,4}$, J. Ses\'e$^{2,4}$, R. C\'ordoba$^{2,4}$, J.M. De Teresa$^{3,4}$ and M.R. Ibarra$^{2,3,4}$}
\address{$^1$ Laboratorio de Bajas Temperaturas, Departamento de F\'isica de la Materia Condensada \\ Instituto de Ciencia de Materiales Nicol\'as Cabrera, Facultad de Ciencias \\ Universidad Aut\'onoma de Madrid, E-28049 Madrid, Spain}
\address{$^2$ Instituto de Nanociencia de Arag\'on, Universidad de Zaragoza, Zaragoza, 50009, Spain}
\address{$^3$ Instituto de Ciencia de Materiales de Arag\'on, Universidad de Zaragoza-CSIC, Facultad de Ciencias, Zaragoza, 50009, Spain}
\address{$^4$ Departamento de F\'isica de la Materia Condensada, Universidad de Zaragoza, 50009 Zaragoza, Spain}

\ead{\mailto{hermann.suderow@uam.es}, \mailto{deteresa@unizar.es}}

\begin{abstract}
We present very low temperature Scanning Tunneling Microscopy and Spectroscopy (STM/S) measurements of a W based amorphous thin film grown with focused-ion-beam. In particular, we address the superconducting properties close to the border, where the thickness of the superconducting film decreases, and the Au substrate emerges. When approaching the Au substrate, the superconducting tunneling conductance strongly increases around the Fermi level, and the quasiparticle peaks do not significantly change its position. Under magnetic fields, the vortex lattice is observed, with vortices positioned very close to the Au substrate.
\end{abstract}

In a recent experiment we have shown that W based thin films grown using a precusor gas and a focused ion beam on top of a conducting Au substrate are excellent superconducting systems for making STM/S down to atomic level and at very low temperatures\cite{Guillamon08c}. The superconducting properties are very homogeneous, and the vortex lattice can be observed with beautiful detail. It is of interest to understand the behavior of the superconductor (S) close to its border, where the height of the thin film continuously decreases, and the normal substrate (N) appears. This gives a curious N-S hybrid system, where the superconducting properties must gradually disappear due to the proximity of the N metal, and where it is possible to follow the positions of the vortices close to the border of the thin film, when the magnetic field is applied perpendicular to the thin film. Note that the situation is exactly the inverse as in Refs.\cite{Truscott99,Vinet01,Gupta04}, where an Au layer was deposited on top of a superconductor, and the local density of states (LDOS) on the N layer measured as a function of its thickness. In those cases, the vortex lattice was not studied.

\begin{figure}[h]
\begin{minipage}{14pc}
\includegraphics[angle=270,width=14pc]{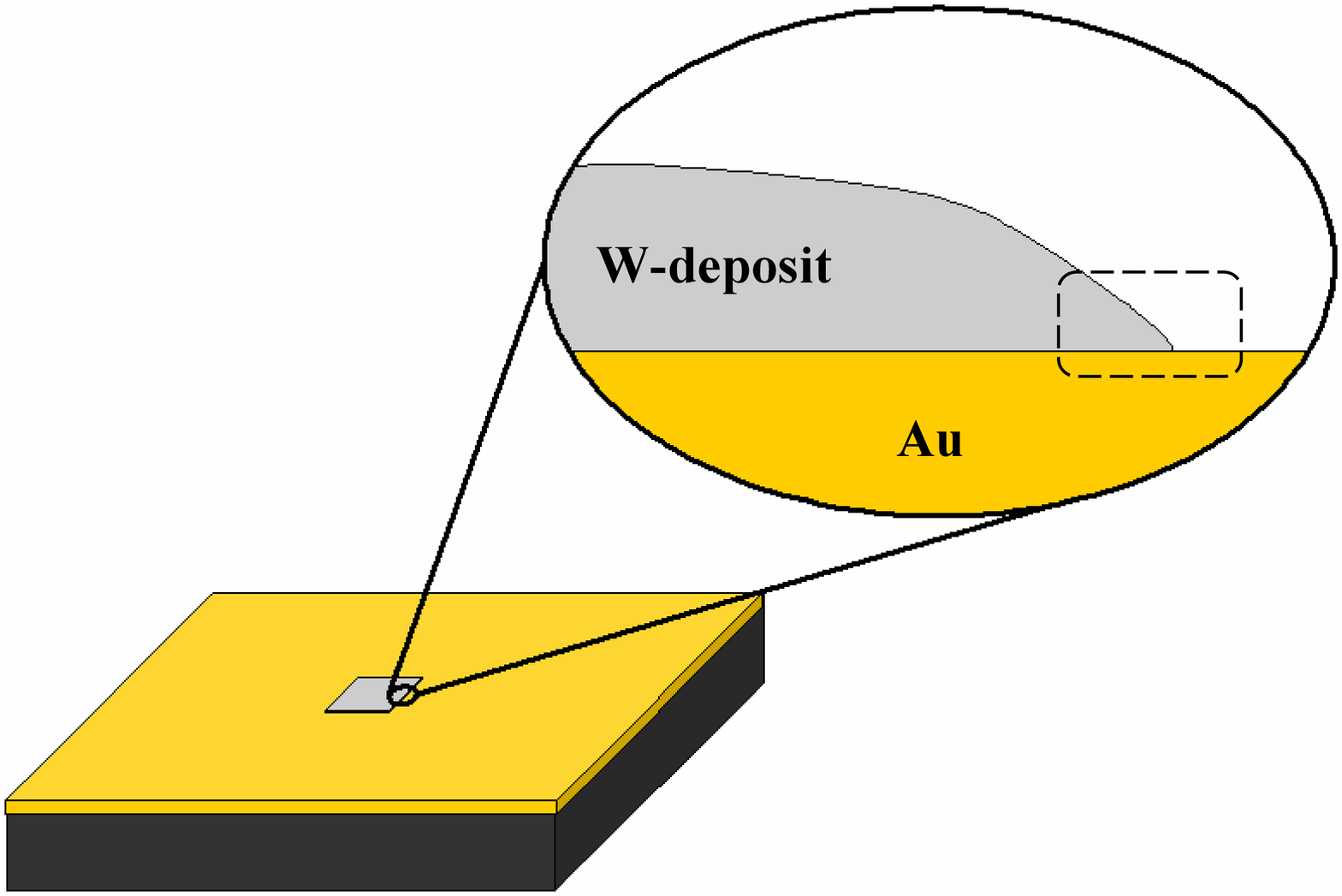}
\caption{Schematic representation of the experiment. We measure the border of a thin film of 30 $\times$ 30 $\mu$m size and 200 nm thickness (figure). A height profile is schematically shown in the zoom up of the figure, where we highlight the area of interest here by a dashed square, which covers the border region between the ion beam grown W based superconducting thin film and Au.\label{Fig0}}
\end{minipage}\hspace{2pc}
\begin{minipage}{20pc}
\includegraphics[angle=270,width=20pc]{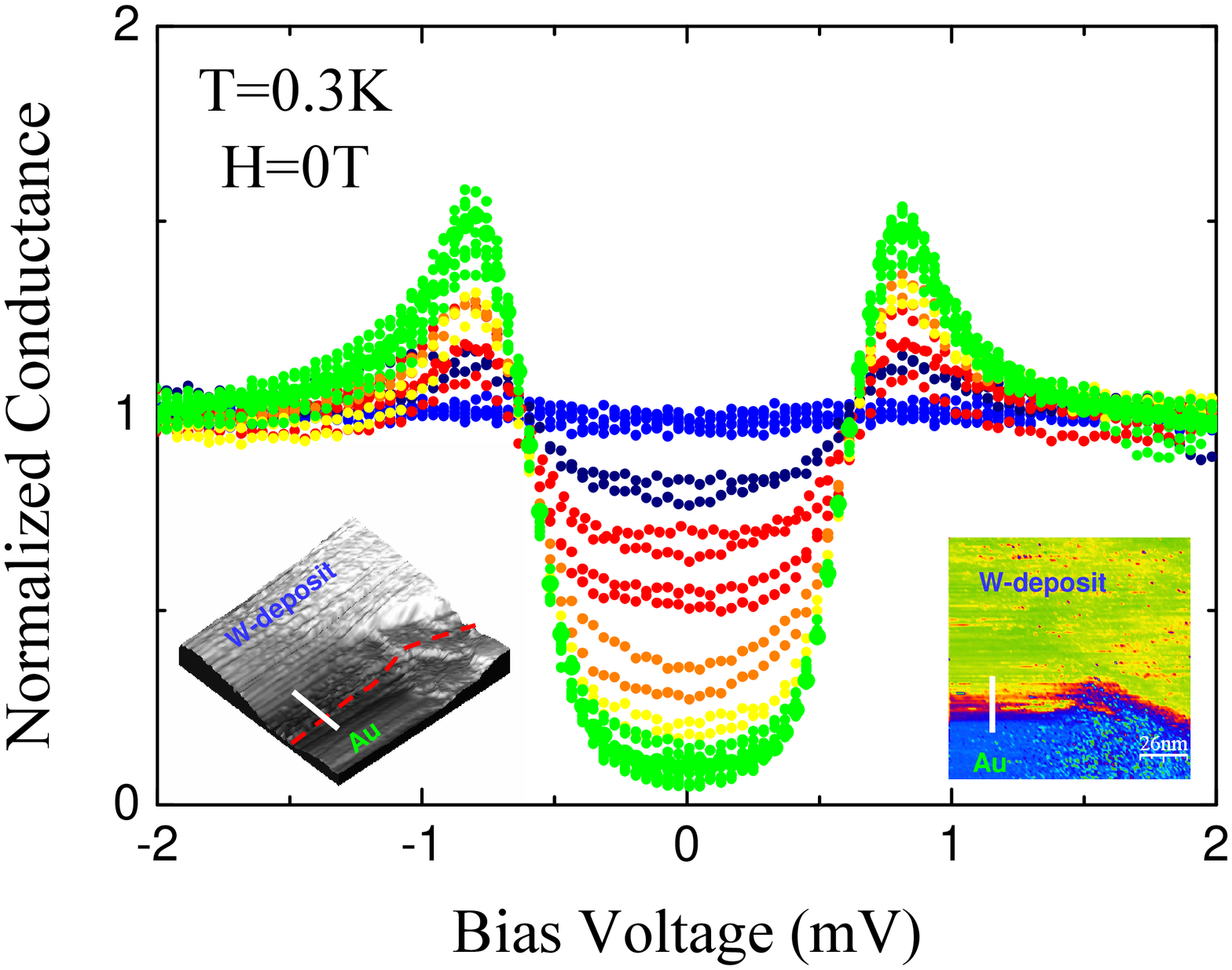}
\caption{Tunneling density of states as a function of the position along a 30 nm line crossing the border of the thin film (curves taken each nm). The total height difference is of about 15 nm, and the tip follows the path highlighted by a white line in the panels. In the left panel we show the topography (as a 3D image), and in the right panel the STS image of the same area obtained from the Fermi level conductance (color scale in correspondence with the figure).\label{Fig1}}
\end{minipage}
\end{figure}

The thin films are grown as described elsewhere, have a concentration of W=40$\pm$7\%, C=43$\pm$4\%, Ga=10$\pm$3\% and O=7$\pm$2\%, a thickness of 200 nm, and T$_c$ = 4.15 K \cite{Guillamon08c,Sadki04,Spoddig07}. Measurements are made in a STM/S system in a dilution refrigerator, which cools sample and tip down to about 100 mK\cite{Guillamon08c,Suderow04,Crespo06a}. We position the tip at the border of the W deposit using an in-situ displacement mechanism. In Fig.\ref{Fig0} we schematically describe our experiment. We focus on the border of the thin film, where topography is smooth, as well within the film \cite{Guillamon08c}, and there is a continuous height decrease towards Au that disappears when reaching the Au surface, which is essentially flat (apart the characteristic surface rugosity of thin films of Au, which are usually granular).

The tunneling spectroscopy as a function of the position at the border of the thin film is shown in Fig.\ref{Fig1}. The superconducting gap is well developed within the deposit, and, around 20 nm from the border to Au, the tunneling characteristics begin to show a significant increase of the low energy conductance, and become flat on top of the Au sample. The position in energy of the quasiparticle peaks does not change when approaching the Au substrate.

It is useful to compare this result with previous work on related structures. Measurements on a normal layer in close contact with a superconductor (as Al or Nb) have shown the appearance of a minigap in the excitation spectrum when the characteristic dimensions of the normal layer are below the electron phase coherence length \cite{Truscott99,Vinet01,Gupta04,Pannetier00,Gueron96,LeSueur08}. In other cases, the continuous reduction of the superconducting correlations due to the proximity to a normal metal results in a decrease of the position in energy of the quasiparticle peaks in the LDOS and a strong increase in the low energy LDOS\cite{Suderow00b,Suderow00c,Suderow02,Rodrigo04b}. However, the behavior observed here is significantly different, as the position of the quasiparticle peaks does not appear to move when approaching the Au sample. Such a behavior cannot be understood in terms of conventional quasiclassical theory\cite{Belzig96}, which reproduces well previously mentioned experiments. Nevertheless, the geometry of our structure is rather 3D, with the superconductor becoming more and more thin when the tip approaches the Au substrate. This may lead to effects not taken into account in quasiclassical theory, which is usually simplified to one dimension. Moreover, we are measuring an amorphous superconducting system, with a strongly reduced mean free path. Indeed, we should note that similar LDOS variations have been observed in previous experiments in disordered TiN films, with highly inhomogeneous superconducting properties\cite{Escoffier04}. It appears that in TiN superconducting "grains" are separated by normal regions, of nominally the same material\cite{Escoffier04}, an effect that has been associated to the closeness to a transition towards an insulating state when further increasing disorder in those systems\cite{Escoffier04,Vinokur08}. It is clear that the W ion beam deposited thin film is a homogeneous superconductor\cite{Guillamon08c}, whereas the TiN films are strongly inhomogeneous. In both cases the mean free path $\ell$ is strongly reduced, and not far to the localization limit (located at $\ell k_F \approx 1$). So the observed LDOS variation may be related to an extremely low mean free path and could be a characteristic feature of N-S hybrid systems made of amorphous S.

\begin{figure}
\includegraphics[width=14cm]{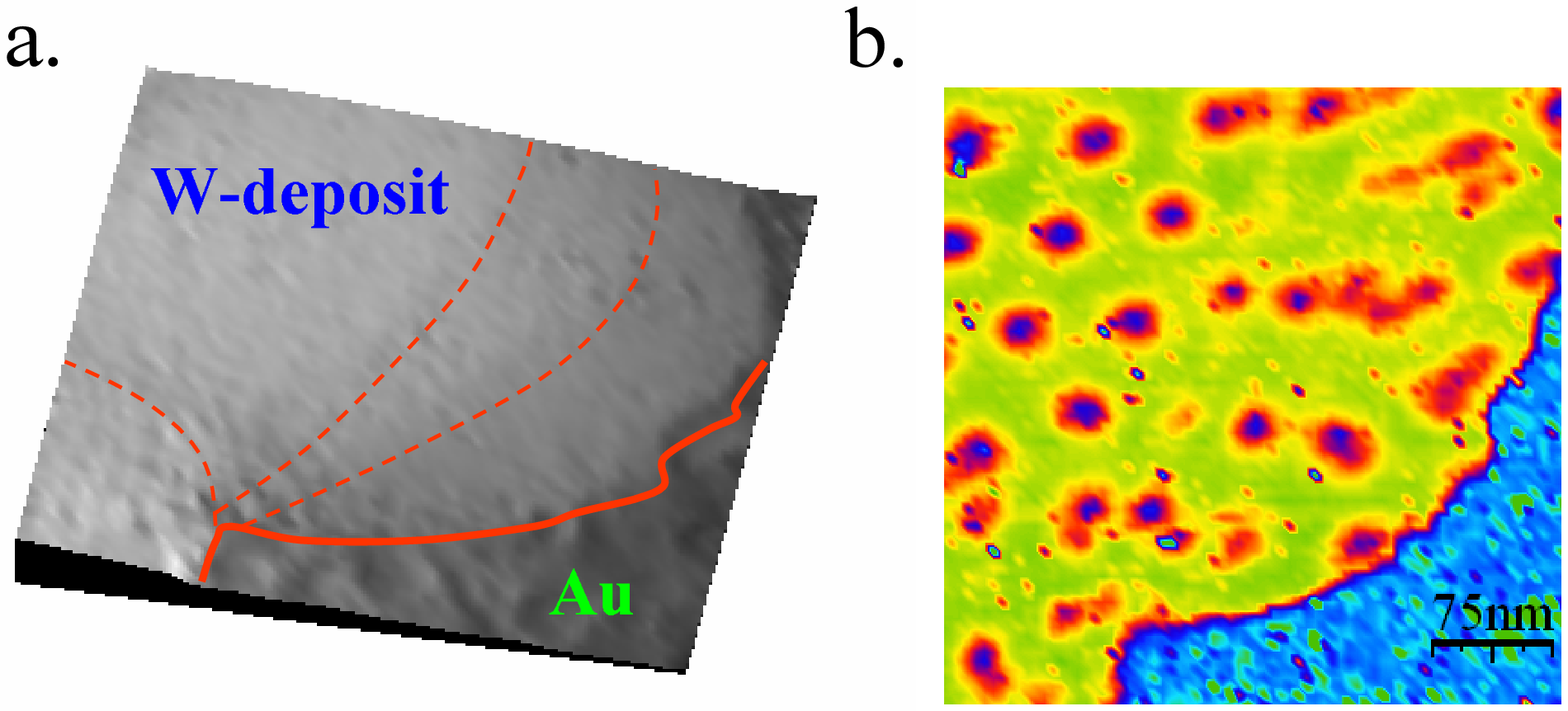}
\caption{In a we show the topography of a typical region at the border of the W ion beam deposited film, as a 3D image, with total height difference of 20 nm. Line highlights the position where the W thin film disappears and Au emerges. Dashed lines show small linear depressions in the topography, seen in height scans, and discussed in Ref.\protect\cite{Guillamon08c}. In b we show the STS image in the same region. Both images have been taken simultaneously at 0.1 K and at 0.7 T.\label{Fig2}}
\end{figure}

In Fig.\ref{Fig2}b we show a STS image made of the Fermi level conductance as a function of the position, and in Fig.\ref{Fig2}a the STM topography of the same area. As shown in Ref.\cite{Guillamon08c}, the orientation of the vortex lattice follows the small linear surface depressions appearing in the topography. The dashed lines in Fig.\ref{Fig2}a highlight the position of these features in the topography, which can be precisely identified in height profiles (not shown). At the upper left corner of the STS image, we observe a hexagon whose orientation is determined by the linear surface depressions. When approaching the border, the vortex lattice becomes disordered, and many elongated patches appear in the STS image. These patches possibly signal vortex bunches or distorted vortex lines. Exactly at the border, in between vortices, we observe qualitatively the same LDOS behavior as in zero field (Fig.\ref{Fig1}), although with additional magnetic field induced pair breaking effects.

Remarkably, in some places we observe vortices which are positioned very close to the Au substrate. This occurs independently of the field history, as well in zero field cooled as in field cooled STS images. In the lower right part of Fig.\ref{Fig2}b we see a single vortex located at about 15 nm from Au. Note that we observe a clear decrease of the Fermi level density of states when leaving the vortex core towards the border, as shown by the color scale in Fig.\ref{Fig2}. However, the distance between the vortex core and the border is much smaller than the intervortex distance at this field (about 60 nm). The reason for having a vortex so close to the border must lie in a delicate balance between the different interactions playing a role in the vortex position. On the one hand, the magnetic field is relatively large (0.7 T), so that the vortex density is high and the energy gained by the vortex when being positioned at the border is small \cite{Buzdin94}. On the other hand, this gain in energy can be compensated by a reduced vortex length due to the decrease in the thickness of the deposit towards Au.

Having a vortex so close to the Au substrate also implies that there are significant dissipation free screening currents circulating in Au. The geometry of the screening currents is determined by the London penetration depth $\lambda$, which is as high as 500 nm in this material\cite{Sadki04,Spoddig07}. The phase coherent length in Au is indeed expected to be pretty large at low temperatures \cite{Pannetier00,Gueron96}. We have increased the temperature close to T$_c$ (=4.15 K here), and have observed no significant variations in the position of the vortices close to the border, showing that the decrease in the phase coherent length must remain above $\lambda$ up to T$_c$, or that the screening currents make up only a small part of the energy balance that positions the vortex close to the border.

In summary, we have followed the superconducting LDOS at the border of an amorphous W ion beam deposited thin film on top of an Au substrate. Remarkably, we observe vortices very close to the border of the W thin film.

\ack

We are indebted to A. Mel'nikov, A.I. Buzdin, F. Guinea, J.G. Rodrigo, V. Crespo, J.P. Brison and J. Flouquet for discussions. The Laboratorio de Bajas Temperaturas is associated to the ICMM of the CSIC. This work was supported by the Spanish MEC (Consolider Ingenio 2010, MAT and FIS programs), by the Comunidad de Madrid through program "Science and Technology in the Millikelvin", and by NES and ECOM programs of the ESF. Images have been rendered using \protect\cite{LaPunta}. \emph{Note added in proof}: The effect of Andreev interference on the LDOS profiles in and around vortices as a consequence of a confined geometry, has been analyzed very recently\cite{Melnikov09}. It may produce a significant shift between the actual vortex center and the maximum in the LDOS. Within this scenario, actual vortex center would lie farther from the border, but the LDOS would be peaked close to the border.

\section*{References}

\providecommand{\newblock}{}

\end{document}